\documentclass[aps,10pt,twocolumn,showpacs,preprintnumbers,amsmath,amssymb,prl]{revtex4-1}

\usepackage{color}
\usepackage{graphicx}
\usepackage{cancel}

\newcommand{\tpsi}{\tilde{\psi}}
\newcommand{\ttrans}{\tilde{\mathcal T}}
\newcommand{\trans}{\mathcal T}
\newcommand{\tG}{\tilde{G}}
\newcommand{\B}{B}
\newcommand{\avT}{\langle T \rangle}
\newcommand{\dif}{d}
\newcommand{\abs}[1]{\left|#1\right|}
\renewcommand{\t}{\gamma}

\newcommand{\lloc}{\ell}

\begin{document}

\title{Point contacts and localization in generic helical liquids}

\author{Christoph~P.~Orth}
\author{Gr\'egory~Str\"ubi}
\author{Thomas~L.~Schmidt}
\affiliation{Department of Physics, University of Basel, Klingelbergstrasse 82, CH-4056 Basel, Switzerland}

\date{\today}

\begin{abstract}
    We consider two helical liquids on opposite edges of a two-dimensional topological insulator, which are connected by one or several local tunnel junctions. In the presence of spatially inhomogeneous Rashba spin-orbit coupling, the spin of the helical edge states is momentum dependent, and this spin texture can be different on opposite edges. We demonstrate that this has a strong impact on the electron transport between the edges. In particular, in the case of many random tunnel contacts, the localization length depends strongly on the spin textures of the edge states.
\end{abstract}

\pacs{71.10.Pm, 72.10.Fk}

\maketitle

\section{I. INTRODUCTION}
An important feature of a two-dimensional topological insulator (quantum spin Hall insulator) is the presence of edge states at the interfaces with a trivial insulator or the vacuum \cite{qi11,hasan10}. These one-dimensional edge states are helical, i.e., the spin of an electron is correlated with its momentum. If the system is time-reversal invariant, two electrons with opposite momenta form a Kramers pair and cannot be coupled by time-reversal invariant perturbations \cite{kane05b,bernevig06b}. As a consequence, non-magnetic impurities should not lead to localization of the electrons propagating on the edges, and the transport is expected to remain ballistic even in the presence of disorder.

Several physical mechanisms lead to deviations from these simple predictions: it was realized early on that interactions can cause inelastic two-particle backscattering, which is allowed by time-reversal invariance, and thus change the conductance at finite temperatures \cite{kane05}. Strong interactions can even open a gap in the edge state spectrum \cite{xu06,wu06}. Moreover, magnetic perturbations can lead to backscattering, and can thus affect the conductance \cite{maciejko09}, or even cause localization \cite{delplace12}.

If we consider a long and narrow two-dimensional topological insulator (length $L$, width $W\ll L$), backscattering between helical states on opposite edges may become possible without breaking time-reversal invariance. However, if $W$ is larger than the decay length of the edge states into the bulk the overlap between states on opposite edges is still exponentially suppressed \cite{zhou08}.
One way to produce backscattering is to couple the helical states on opposite edges by local tunneling \cite{strom09,hou09,teo09}. On the one hand, such processes may be realized intentionally at point contacts formed either by lithographic techniques or by appropriate gating \cite{liu08}, depending on the topological insulator material, and interesting transport properties have been predicted \cite{liu11,schmidt11,dolcini11,lee12,dolcetto12,edge13}. On the other hand, tunneling between opposite edges may also emerge accidentally in narrow samples if the bulk material is sufficiently disordered, so that charge puddles \cite{skinner12,vayrynen13} can connect opposite edges. As long as this transport remains elastic and the puddles are dilute, such a system can be modeled using local tunnel contacts at random positions and with random tunnel amplitudes.

To investigate these systems, we are going to employ the concept of \emph{generic} helical liquids proposed in Ref.~\cite{schmidt12}. Such a generic helical liquid provides a rather general template for time-reversal invariant helical edge states in which the electron spin is not necessarily a good quantum number. This situation can arise in all proposed topological insulator materials to date by effects such as bulk inversion asymmetry, structural inversion asymmetry \cite{qi11}, or Rashba spin-orbit coupling \cite{rothe10}. The most important consequence of the broken axial spin symmetry is a rotation of the spin quantization axis of the helical eigenstates as a function of momentum.

It was shown in Ref.~\cite{schmidt12} that in this case, the projections of left-moving and right-moving eigenmodes on a fixed spin axis are determined by a momentum-dependent spin rotation matrix $\B_k$. Certain symmetries of the rotation matrix $\B_k$, which we refer to as the ``spin texture'' of the edge state, are fixed by unitarity and time-reversal invariance, but its amplitude can be tuned locally: for instance, in HgTe quantum wells, a spatially inhomogeneous electric field perpendicular to the plane of the topological insulator will induce Rashba spin-orbit coupling \cite{rothe10} with different amplitudes on different sample edges. In such an experiment, the spins of the helical states at the Fermi energy $\mu$ on opposite edges will be tilted relative to each other by an angle $\theta(\mu)$. In this paper, we will show that a nonzero $\theta(\mu)$ has a strong effect on the current-voltage characteristic.

\begin{figure}[t]
  \includegraphics[width=\columnwidth]{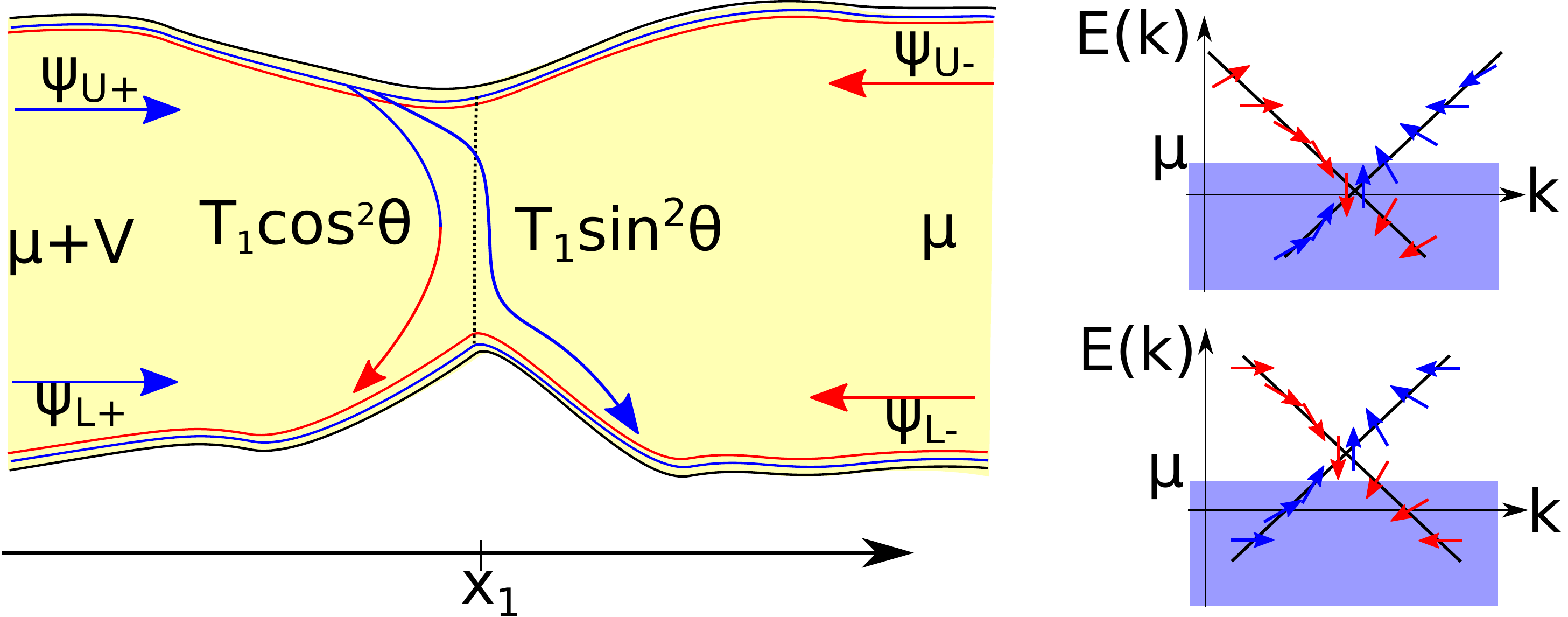}
  \caption{For a single point contact, the angle $\theta(\mu)$, see Eq.~(\ref{eq:thetamu}), determines the branching ratio of right movers on the upper edge into right- and left-movers on the lower edge. The right panel shows the spectra of the upper and lower edges for $V=0$. $\theta(\mu)$ is given by the difference in spin axis rotation at the Fermi energy. For clarity, we set $\vartheta_1=0$, which corresponds to spin-conserving tunneling.}
  \label{fg:SinglePointContact}
\end{figure}

Our starting point is two generic helical liquids with different spin rotation matrices $B_{k,U}$ and $B_{k,L}$ living on the upper and lower edges, respectively, of a narrow 2D topological insulator. We will calculate two-terminal transport properties, where a bias voltage is applied between the left side and the right side of the sample as shown in Fig.~\ref{fg:SinglePointContact}. First, we will investigate the effect of a single tunnel contact between the upper and lower edges, and show that tunneling can lead to forward scattering as well as backscattering depending on $\theta(\mu)$, see Fig.~\ref{fg:SinglePointContact}. Next, we will add a second tunnel junction and show that interference effects make it possible to determine $\theta(\mu)$ from a conductance measurement. Finally, we will consider a large number of random tunnel contacts. Even for conventional helical liquids where spin is conserved, this type of disorder leads to localization and to a suppression of the conductance $G \propto 2 G_0 e^{-L/\lloc}$, where $G_0 = e^2/h$ is the conductance quantum, $L$ is the sample length and $\lloc$ is the localization length. For our model, we shall show that $\lloc$ depends strongly on $\theta(\mu)$. Since $\theta(\mu)$ depends on the Rashba spin-orbit coupling strength, we predict that the localization length of a narrow two-dimensional topological insulator in a two-terminal configuration is strongly sensitive to a spatially inhomogeneous electric field.

\section{II. ONE POINT CONTACT}
We start by considering a single point contact between two helical edges along the $x$ direction. After linearizing the spectrum, the kinetic part of the Hamiltonian for the two edges is given by
\begin{align}
 H_{kin}&=-i v_F \sum_{\eta=U,L} \sum_{\alpha=\pm} \alpha \int_{-\infty}^\infty \dif x \, \psi_{\eta \alpha}^\dag(x) \partial_x \psi_{\eta \alpha}(x),
\end{align}
where $\eta$ labels the upper ($\eta=U$) and lower ($\eta=L$) edges, each of which hosts right-movers ($\alpha=+$) and left-movers ($\alpha=-$). We use units $\hbar=1, e=1$. In general, the eigenstates $\psi_{k,\eta\alpha}$ of this Hamiltonian, where $k$ is the momentum along the $x$ direction, need not be spin eigenstates \cite{schmidt12}: effects such as bulk inversion asymmetry, structural inversion asymmetry or Rashba spin-orbit coupling break the axial spin symmetry. The projections of the helical eigenstates $\psi_{k,\eta\alpha}$ on states with definite spin along an arbitrary but fixed spin quantization axis, $\tpsi_{k,\eta\sigma}$ with $\sigma \in \{\uparrow,\downarrow\}$, can be encoded into a $2\times 2$ momentum-dependent rotation matrix \cite{schmidt12}
\begin{align} \label{eq:tpsiBpsi}
 \left( \begin{matrix} \tpsi_{k,\eta\uparrow} \\ \tpsi_{k,\eta\downarrow} \end{matrix} \right) = \B_{k,\eta} \left( \begin{matrix} \psi_{k,\eta +} \\ \psi_{k,\eta -} \end{matrix} \right) ,
\end{align}
where $\eta = U,L$. The two rotation matrices $\B_{k,\eta}$ of the upper and lower edge are in principle independent of each other because the spatially separated edges can be subject to different gate voltages and Rashba spin-orbit coupling strengths. The resulting band structure is schematically shown in Fig.~\ref{fg:SinglePointContact}. The matrices $\B_{k,\eta}$ are unitary, and as a consequence of time-reversal invariance, satisfy the condition $\B_{k,\eta}= \B_{-k,\eta}$. Concrete expressions for $\B_{k,\eta}$ for helical edge states in HgTe/CdTe quantum wells in the presence of Rashba spin-orbit coupling where shown in Ref.~\cite{schmidt12}.

As a first step, we are going to consider the effect of a single point contact at the position $x_1$ between the two helical edges, see Fig.~\ref{fg:SinglePointContact}. The Hamiltonian is $H = H_{kin} + H_{1}(x_1)$, where
\begin{align}\label{eq:HT}
  H_{j}(x) = \t_j \cos(\vartheta_j) &\sum_{\sigma=\uparrow\downarrow} \left[ \tpsi_{U \sigma}^\dag(x) \tpsi_{L \sigma}(x) + \text{H.c.} \right] \notag \\
  + \t_j \sin(\vartheta_j) &\sum_{\sigma=\uparrow\downarrow} \left[ \tpsi_{U \sigma}^\dag(x) \tpsi_{L -\sigma}(x) + \text{H.c.} \right]. 
\end{align}
We allow both spin-conserving and spin-flip tunneling, parametrized with the real amplitudes $\t_1 \cos(\vartheta_1)$ and $\t_1 \sin(\vartheta_1)$. The Fermi wavelength should be large compared to the width of the tunneling region in order to model the tunneling as local.

Our first goal is to determine the total current flowing along the $x$-direction if the system is coupled to two reservoirs held at chemical potentials $\mu_-=\mu$ and $\mu_+=\mu+V$ on the right and left sides, respectively. These reservoirs thus define the chemical potential of the right-movers ($\mu + V$) and left-movers ($\mu$). For the calculation of the current in this nonequilibrium situation, we employ Keldysh Green's functions (GF), which we define as (for $\alpha \in \{+,-\}$, $\sigma,\tau \in \{ \uparrow,\downarrow\}$ and $\eta \in \{ U, L\}$)
\begin{align}
 g_{\eta}^{\alpha \alpha}(x,t) &= -i \langle T_C \psi_{\eta\alpha}(x,t) \psi^\dag_{\eta\alpha}(0,0) \rangle_{0} \\
 \tilde{g}_{\eta}^{\sigma \tau}(x,t) &= -i \langle T_C \tilde{\psi}_{\eta\sigma}(x,t) \tilde{\psi}^\dag_{\eta\tau}(0,0) \rangle_{0} .
\end{align}
Here, $T_C$ denotes the Keldysh time-ordering operator and the expectation value is taken with respect to the unperturbed Hamiltonian $H_{kin}$. The Fourier transforms of these GF are related via the spin-rotation matrices,
\begin{align}
 \tilde{g}_{\eta}^{\sigma \tau}&(k,\omega)  = \sum_{\alpha=\pm} B_{k,\eta}^{\sigma \alpha} \left[ B_{k,\eta}^{-1}\right]^{\alpha \tau} g_{\eta}^{\alpha \alpha}(k,\omega) ,
\end{align}
where the unperturbed Green's function of the eigenstates of $H_{kin}$ at zero temperature reads
\begin{align}
 & g^{\alpha \alpha}_{\eta}(k,\omega) = \notag \\ &\left( \begin{matrix} (\omega-\alpha v_F k + i s \delta)^{-1} & - i \pi \delta(\omega-\alpha v_F k) \left( s-1\right)  \\  - i \pi \delta(\omega-\alpha v_F k)\left( s+1\right) & - (\omega - \alpha v_F k- i s \delta)^{-1} \end{matrix}\right) .
\end{align}
Here $\delta=0^+$ is infinitesimal and $s=\text{sign}(\omega-\mu_\alpha)$. To obtain the Green's function in the presence of the single point contact from the unperturbed one, we use a Dyson equation. For the spin-resolved Green's function $\tilde{G}_{\eta \zeta}^{\sigma \tau}(x,x',\omega)$, we find (for $\sigma,\tau \in \{ \uparrow,\downarrow\}$ and $\eta,\zeta \in \{U,L\}$)
\begin{align} \label{eq:DysonEq}
 \tilde{G}_{\eta \zeta}^{\sigma \tau}(x,x',\omega) =& \tilde{g}_\eta^{\sigma \tau}(x-x',\omega)\delta_{\eta \zeta} \notag \\
 + \t_1 \cos(\vartheta_1) &\sum_{\upsilon = \uparrow\downarrow} \tilde{g}_\eta^{\sigma \upsilon} (x-x_1,\omega) \tilde{G}_{-\eta \zeta}^{\upsilon \tau}(x_1,x',\omega)  \notag \\
 + \t_1 \sin(\vartheta_1) &\sum_{\upsilon = \uparrow\downarrow} \tilde{g}_\eta^{\sigma \upsilon} (x-x_1,\omega) \tilde{G}_{-\eta \zeta}^{-\upsilon \tau}(x_1,x',\omega).
\end{align}
In general, the total current can be written as an integral over a function which depends only on the combination $B_{k,U}^\dag B_{k,L}$ of spin rotation matrices. To obtain a simpler expression we use the following form of $\B_{k,\eta}$ \cite{schmidt12}
\begin{equation}\label{eq:B}
 \B_{k,\eta} = \left( \begin{matrix} \cos(\theta_{k,\eta}) & -\sin(\theta_{k,\eta}) \\ \sin(\theta_{k,\eta}) & \cos(\theta_{k,\eta}) \end{matrix} \right) ,
\end{equation}
where, according to Eq.~(\ref{eq:tpsiBpsi}), the angle $\theta_{k,\eta}$, as a function of momentum $k$ and the edge $\eta$, determines the rotation of the spin quantization axis of the eigenstates of $H_{kin}$, with respect to the fixed spin orientations $\uparrow,\downarrow$.
As shown in Ref.~\cite{schmidt12}, one obtains Eq.~(\ref{eq:B}), e.g., by adding Rashba spin-orbit coupling to the Bernevig-Hughes-Zhang Hamiltonian for HgTe/CdTe quantum wells~\cite{bernevig06}, or for topological insulators based on InAs/GaSb heterostructures \cite{liu08,knez11}. For the current, we find $I(V) = \int^V_0 d\omega G(\mu + \omega)$, where $G(\mu)$ is the two-terminal differential conductance for a system held at chemical potential $\mu$,
\begin{align}\label{eq:G1}
    G(\mu) = 2 G_0 \left\{1- T_1 \cos^2[\theta(\mu)+\vartheta_1] \right\}.
\end{align}
Here, $G_0=1/(2\pi)=e^2/h$ is the conductance quantum and $T_1=4 \Gamma_1/(1+\Gamma_1)^2$ the tunnel probability with $\Gamma_1 = \t_1^2/(2 v_F)^2$. 
The conductance and the current depend only on the difference
\begin{equation}\label{eq:thetamu}
 \theta(\mu) = \theta_{\mu/v_F,U}-\theta_{\mu/v_F,L} ,
\end{equation}
where the angles $\theta_{\mu/v_F,\eta}$ for $\eta=U,L$ are defined by Eq.~(\ref{eq:B}). The conductance $G(\mu)$ is a function of the angle $\vartheta_1$, which depends on the microscopic details of the tunnel junction, as well as on $\theta(\mu)$, which is a global property of the helical liquid.
According to Eq.~(\ref{eq:G1}), the angle difference $\theta(\mu)$ has a strong impact on the conductance, because it determines the amount of right-movers on the upper edge that become left-movers or right-movers in the lower edge after tunneling, see Fig.~\ref{fg:SinglePointContact}. In the limit of strong tunneling ($T_1\to1$), the conductance from left to right can change from zero at $\theta(\mu) = -\vartheta_1$ to $2 G_0$ at $\theta(\mu) = \pi/2-\vartheta_1$.
\begin{figure}[t]
  \includegraphics[width=\columnwidth]{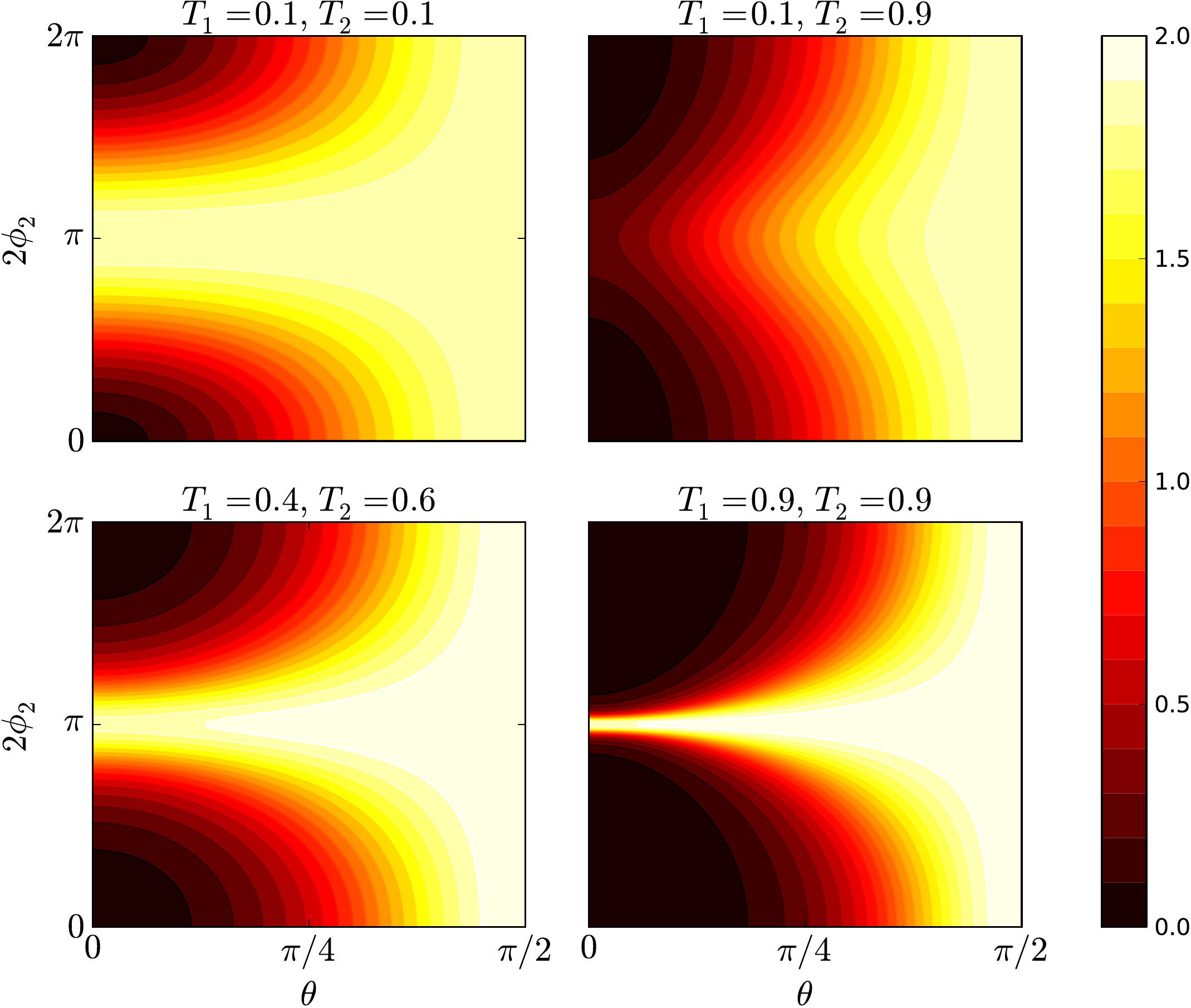}
  \caption{Conductance (in units of $e^2/h$) of a 2D topological insulator with two tunnel junctions as a function of the angle $\theta(\mu)$ and distance between tunnel junctions $\Delta x_2$, where $\phi_2= \Delta x_2 \mu/v_F$. We use $\vartheta_1=\vartheta_2=0$, which corresponds to spin-conserving tunneling.}
  \label{fg:2ContactConductance}
\end{figure}

\section{III. TWO POINT CONTACTS}
Next, we extend the system by adding a second point contact \cite{chu09,dolcini11,virtanen11,romeo12} at position $x_2$ with the tunnel amplitude $\t_2$ and angle $\vartheta_2$, i.e., $H = H_{kin} + H_1(x_1) + H_2(x_2)$.
The distance between the two point contacts is $\Delta x_2 = x_2 - x_1$. From here on, we assume that the rotation angles $\theta_{k,\eta}$ change very slowly with $k$, only on a large momentum scale $k_0 \gg k_F$. Furthermore we assume that the distance between tunnel contacts is large compared to the Fermi wavelength, i.e., $\Delta x_2 \gg 1/k_F$. To calculate the current in this system we have to solve a Dyson equation analogous to Eq.~(\ref{eq:DysonEq}) but with a second contribution for the second tunnel contact at $x_2$.
At the special points $x,x' \in \{x_1,x_2\}$ this Dyson equation forms a system of linear equations which we can solve for $\tG$. In a second step, we use this solution in the Dyson equation again to find also the solution for arbitrary $x,x'$. For real tunneling amplitudes and using Eq.~(\ref{eq:B}), the conductance reads
  \begin{multline} \label{eq:2PointContactCurrent}
   G(\mu) = \\ 
   \frac{2G_0\left\{1- T_1 \cos^2[\theta(\mu)+\vartheta_1] \right\}\left\{1- T_2 \cos^2[\theta(\mu)+\vartheta_2] \right\}}{\abs{1+ \sqrt{T_1T_2} \, \cos\left[\theta(\mu)+\vartheta_1\right] \cos\left[ \theta(\mu)+\vartheta_2\right] \, e^{2 i \Delta x_2 \mu/v_F}}^2}.
  \end{multline}
We find interference patterns between the different paths that depend on the phase $\phi_2 = \Delta x_2\mu/v_F$ acquired by an electron when passing through the loop formed by the two tunnel junctions. Multiple traversals of this loop yield a geometric series which leads to the denominator of Eq.~(\ref{eq:2PointContactCurrent}). The conductance as a function of the phase $\phi_2$ and the spin rotation angle $\theta(\mu)$ is plotted for different tunnel probability combinations in Fig.~\ref{fg:2ContactConductance}. For equal tunnel strengths $T_1=T_2$, we always find the maximal conductance of $2G_0$ at the resonance condition $\phi_2=\pi$, even for very weak tunneling. This can be interpreted as Fabry-P\'erot resonances which remain visible in the current as long as $V\ll v_F/\Delta x_2$. In setups where the distance between the two tunnel contacts is tunable without effecting the other parameters, varying $\Delta x_2$ makes it possible to measure $\theta(\mu)$ as a function of $\mu$.

\section{IV. DISORDERED SYSTEM}
\begin{figure}[t]
  \includegraphics[width=\columnwidth]{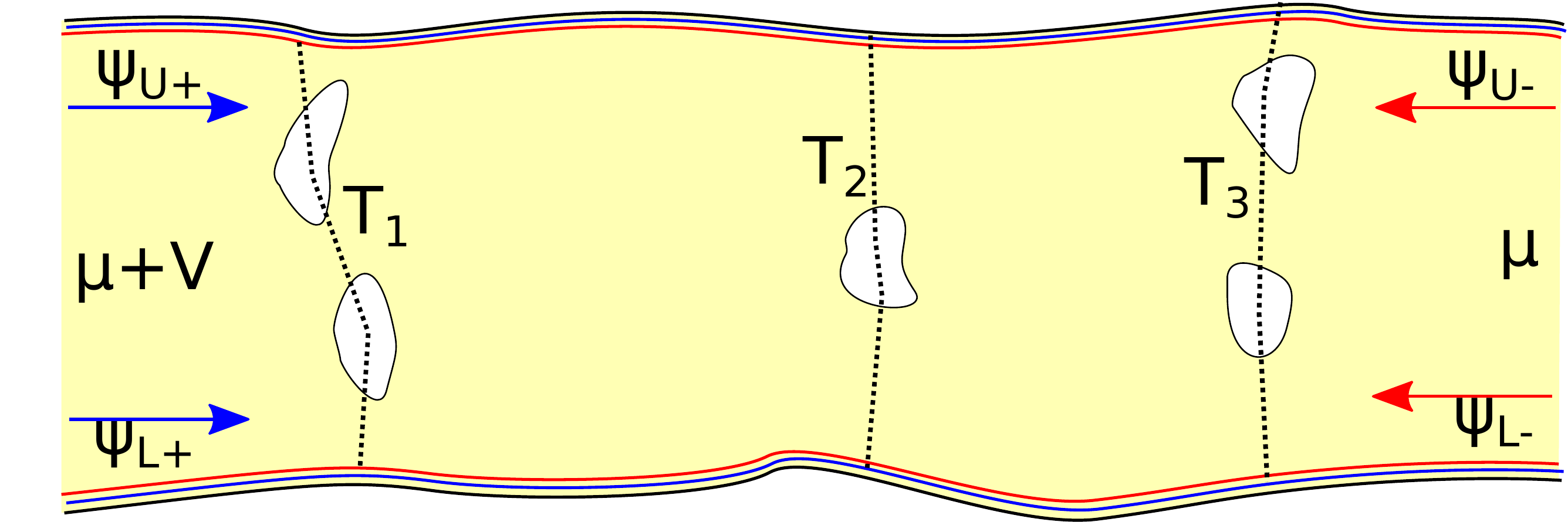}
  \caption{Charge puddles in a narrow 2D topological insulator between two edges can form tunnel paths at random positions $x_j$ and with random tunnel amplitudes $\t_j$. Tunneling leads to localization, with a localization length which depends strongly on the spin rotation angle $\theta(\mu)$. }
  \label{fg:WideDisorderedWire}
\end{figure}

Finally we will examine a large number $N$ of point contacts between the two edges at random positions $x_j$ and with random tunnel amplitudes $\t_j$, $\vartheta_j$ as depicted in Fig.~\ref{fg:WideDisorderedWire}. At the end we will perform a disorder average over the $x_j$, $\t_j$ and $\vartheta_j$. This model captures the physics of narrow samples where the helical edges are connected by charge puddles \cite{skinner12,vayrynen13,koenig13} that originate from doping. We effectively describe this by a set of tunnel Hamiltonians
\begin{equation}
 H_{N} = \sum_{j=1}^N H_j(x_j) ,
\end{equation}
which we add to $H_{kin}$. We assume that the positions $x_j$ are uniformly distributed in a region of length $L$, and that the density $n=N/L$ of point contacts is small, i.e., $k_0 \gg k_F \gg n$.

For the disorder average over many point contacts it is most convenient to use transfer matrices to calculate the current. The transfer matrix $\trans_j$ ($j=1,\dots, N$) relates the states on the left side of the $j$th point contact to those on its right side,
\begin{align}
 \trans_j(\mu) &= \frac{1}{1-T_j \cos^2(\theta+\vartheta_j)}\left( \begin{matrix} A & B \\ -B^* & A^* \end{matrix}\right) \\
 A &= \left( \begin{matrix} \sqrt{1-T_j} & -i \sqrt{T_j} \sin(\theta+\vartheta_j) \\ -i \sqrt{T_j} \sin(\theta+\vartheta_j) & \sqrt{1-T_j} \end{matrix}\right) \nonumber \\
 B &= \cos(\theta+\vartheta_j)\left( \begin{matrix} -T_j \sin(\theta+\vartheta_j) & i \sqrt{(1-T_j)T_j} \\ -i \sqrt{(1-T_j)T_j} & T_j \sin(\theta+\vartheta_j) \end{matrix}\right) , \nonumber
\end{align}
in the basis $(\psi_{U+}, \psi_{L+}, \psi_{U-}, \psi_{L-})$. The total transfer matrix $\ttrans_N$ for transport through $N$ point contacts can be determined from the following recursive relation \cite{delplace12}
\begin{align}\label{eq:ttransN}
 \ttrans_1 = \trans_1,\quad
 \ttrans_{j} = \ttrans_{j-1} P_{j} \trans_{j} ,
\end{align}
where $P_j = \text{diag} \left( e^{i \phi_j} , e^{i \phi_j} , e^{-i \phi_j} , e^{-i \phi_j} \right)$ is a diagonal transfer matrix that describes free propagation between the contacts at positions $x_{j-1}$ and $x_j$, and results in dynamical phases $\phi_j=\Delta x_j \mu/v_F$ for right-movers and $-\phi_j$ for left-movers, where $\Delta x_j = x_j - x_{j-1}$.
The total transfer matrix has a general structure
\begin{align}
 \ttrans_j &= \left( \begin{matrix} \lambda_j & \rho_j \\ - \rho^*_j & \lambda^*_j \end{matrix}\right) \notag \\
  \lambda_j &= \left( \begin{matrix} a_j & i a_j b_j \\ i a_j b_j & a_j \end{matrix}\right), \quad \rho_j = \left( \begin{matrix} c_j & i c_j/b_j \\ -i c_j/b_j & -c_j \end{matrix}\right)
\end{align}
with two complex parameters $a_j,c_j$ and a real parameter $b_j$. Products of these transfer matrices always reproduce this structure. The lower right component of the transfer matrix is the inverse of the transmission matrix $t_j = (\lambda^*_j)^{-1}$ which is connected to the dimensionless conductance through $j$ contacts as $\tG_j(\mu) = \text{Tr}( t^\dag_j t_j )$ \cite{bruus04}. Equation~\ref{eq:ttransN} leads to a recursive relation expressing the conductance through $j$ contacts in terms of the product of the conductance through $j-1$ contacts, the conductance of the $j$th contact, as well as a phase dependent term
\begin{equation}
 \tG_j = \tG_{j-1} \frac{G_j}{2 G_0} \abs{1+\frac{c_{j-1}}{a_{j-1} b_{j-1}} \sqrt{T_j} \cos(\theta+\vartheta_j) e^{-2 i \phi_j}}^{-2}
\end{equation}
where $G_j = 2 G_0 -2 G_0 T_j \cos^2(\theta+\vartheta_j)$ and $\tG_0 = 2$. We take the logarithm of the whole equation and solve it recursively. Averaging both sides over the phases $\phi_j$ yields for the total conductance of all $N$ contacts,
\begin{equation}
 \langle \log(\tilde{G}_N/2) \rangle_{\phi} = \log\left(\frac{G_N}{2G_0}\right) + \dots +\log\left(\frac{G_1}{2G_0}\right).
\end{equation}
Furthermore, taking the average over the tunnel probabilities $T_j$ and angles $\vartheta_j$ renders all the logarithms on the right side equal. Comparing the resulting equation
\begin{equation}
 \langle \log(\tilde{G}_N/2) \rangle = N \left\langle \log\left\{ 1-T \cos^2\left[\theta(\mu)+\vartheta\right]\right\} \right\rangle
\end{equation}
with $\langle \log(\tG_N/2) \rangle = -L/\lloc$ allows us to define the localization length as \cite{delplace12,anderson80,pendry94}
\begin{align}\label{eq:loc}
 \lloc^{-1} =& - \lim_{N\rightarrow \infty} \frac{n}{N} \left\langle \log(\tilde{G}_N/2) \right\rangle \nonumber \\
    =& - n \left\langle \log\left\{ 1-T \cos^2\left[\theta(\mu)+\vartheta\right]\right\} \right\rangle .
\end{align}
If the tunnel probabilities are small ($T_j \approx 0$), spin flip tunneling is suppressed ($\vartheta_j \approx 0$) and $T_j$ and $\vartheta_j$ are statistically independent variables we can approximate the logarithm to linear order and obtain
\begin{equation}
 \lloc^{-1}  =  n \avT \cos^2\left[\theta(\mu)\right].
\end{equation}\
This means a change of $\theta(\mu)$ can change the localization length in the interval $1/(n \avT) < \lloc < \infty$. In particular, an infinite localization length can be reached even in the presence of tunneling for $\theta(\mu)=\pi/2$. 

\section{V. CONCLUSIONS}
To summarize, we have investigated narrow two-dimensional topological insulators where the helical liquids on the two opposite edges have different spin textures. In this case, the spin axes of two particles on the upper and lower edges at the Fermi energy $\mu$ are tilted by an angle $\theta(\mu)$. This angle can be tuned in experiments, e.g., by applying a perpendicular electric field gradient.
We have considered a system where the edge states are coupled locally by one or several tunnel contacts. For two tunnel contacts, interference effects in the two-terminal conductance make it possible to determine $\theta(\mu)$. Many random contacts lead to localization of the edge states with a strong dependence of the localization length on $\theta(\mu)$. These effects could be used in experiments to map the spin-structure of the helical edge states.

\acknowledgments
\section{ACKNOWLEDGMENTS}
We would like to acknowledge stimulating discussions with C. Bruder. This work was financially supported by the Swiss NSF and the NCCR Quantum Science and Technology.

\bibliography{references}

\begin{thebibliography}{10}%
\makeatletter
\providecommand \@ifxundefined [1]{%
 \ifx #1\undefined \expandafter \@firstoftwo
 \else \expandafter \@secondoftwo
\fi
}%
\providecommand \@ifnum [1]{%
 \ifnum #1\expandafter \@firstoftwo
 \else \expandafter \@secondoftwo
\fi
}%
\providecommand \enquote [1]{``#1''}%
\providecommand \bibnamefont  [1]{#1}%
\providecommand \bibfnamefont [1]{#1}%
\providecommand \citenamefont [1]{#1}%
\providecommand\href[0]{\@sanitize\@href}%
\providecommand\@href[1]{\endgroup\@@startlink{#1}\endgroup\@@href}%
\providecommand\@@href[1]{#1\@@endlink}%
\providecommand \@sanitize [0]{\begingroup\catcode`\&12\catcode`\#12\relax}%
\@ifxundefined \pdfoutput {\@firstoftwo}{%
 \@ifnum{\z@=\pdfoutput}{\@firstoftwo}{\@secondoftwo}%
}{%
 \providecommand\@@startlink[1]{\leavevmode\special{html:<a href="#1">}}%
 \providecommand\@@endlink[0]{\special{html:</a>}}%
}{%
 \providecommand\@@startlink[1]{%
  \leavevmode
  \pdfstartlink
   attr{/Border[0 0 1 ]/H/I/C[0 1 1]}%
   user{/Subtype/Link/A<</Type/Action/S/URI/URI(#1)>>}%
  \relax
 }%
 \providecommand\@@endlink[0]{\pdfendlink}%
}%
\providecommand \url  [0]{\begingroup\@sanitize \@url }%
\providecommand \@url [1]{\endgroup\@href {#1}{\urlprefix}}%
\providecommand \urlprefix [0]{URL }%
\providecommand \Eprint[0]{\href }%
\@ifxundefined \urlstyle {%
  \providecommand \doi [1]{doi:\discretionary{}{}{}#1}%
}{%
  \providecommand \doi [0]{doi:\discretionary{}{}{}\begingroup
  \urlstyle{rm}\Url }%
}%
\providecommand \doibase [0]{http://dx.doi.org/}%
\providecommand \Doi[1]{\href{\doibase#1}}%
\providecommand \bibAnnote [3]{%
  \BibitemShut{#1}%
  \begin{quotation}\noindent
    \textsc{Key:}\ #2\\\textsc{Annotation:}\ #3%
  \end{quotation}%
}%
\providecommand \bibAnnoteFile [2]{%
  \IfFileExists{#2}{\bibAnnote {#1} {#2} {\input{#2}}}{}%
}%
\providecommand \typeout [0]{\immediate \write \m@ne }%
\providecommand \selectlanguage [0]{\@gobble}%
\providecommand \bibinfo [0]{\@secondoftwo}%
\providecommand \bibfield [0]{\@secondoftwo}%
\providecommand \translation [1]{[#1]}%
\providecommand \BibitemOpen[0]{}%
\providecommand \bibitemStop [0]{}%
\providecommand \bibitemNoStop [0]{.\EOS\space}%
\providecommand \EOS [0]{\spacefactor3000\relax}%
\providecommand \BibitemShut [1]{\csname bibitem#1\endcsname}%
\bibitem{qi11}%
  \BibitemOpen
  \bibfield{author}{%
  \bibinfo {author} {\bibfnamefont{X.}~\bibnamefont{Qi}}\ and\ \bibinfo
  {author} {\bibfnamefont{S.}~\bibnamefont{Zhang}},\ }%
  \bibfield{journal}{%
  \bibinfo {journal} {Rev. Mod. Phys.}\ }%
  \textbf{\bibinfo {volume} {83}},\ \bibinfo {pages} {1057} (\bibinfo {year}
  {2011})%
  \bibAnnoteFile{NoStop}{qi11}%
\bibitem{hasan10}%
  \BibitemOpen
  \bibfield{author}{%
  \bibinfo {author} {\bibfnamefont{M.~Z.}\ \bibnamefont{Hasan}}\ and\ \bibinfo
  {author} {\bibfnamefont{C.~L.}\ \bibnamefont{Kane}},\ }%
  \bibfield{journal}{%
  \bibinfo {journal} {Rev. Mod. Phys.}\ }%
  \textbf{\bibinfo {volume} {82}},\ \bibinfo {pages} {3045} (\bibinfo {year}
  {2010})%
  \bibAnnoteFile{NoStop}{hasan10}%
\bibitem{kane05b}%
  \BibitemOpen
  \bibfield{author}{%
  \bibinfo {author} {\bibfnamefont{C.~L.}\ \bibnamefont{Kane}}\ and\ \bibinfo
  {author} {\bibfnamefont{E.~J.}\ \bibnamefont{Mele}},\ }%
  \bibfield{journal}{%
  \bibinfo {journal} {Phys. Rev. Lett.}\ }%
  \textbf{\bibinfo {volume} {95}},\ \bibinfo {pages} {146802} (\bibinfo {year}
  {2005})%
  \bibAnnoteFile{NoStop}{kane05b}%
\bibitem{bernevig06b}%
  \BibitemOpen
  \bibfield{author}{%
  \bibinfo {author} {\bibfnamefont{B.~A.}\ \bibnamefont{Bernevig}}\ and\
  \bibinfo {author} {\bibfnamefont{S.}~\bibnamefont{Zhang}},\ }%
  \bibfield{journal}{%
  \bibinfo {journal} {Phys. Rev. Lett.}\ }%
  \textbf{\bibinfo {volume} {96}},\ \bibinfo {pages} {106802} (\bibinfo {year}
  {2006})%
  \bibAnnoteFile{NoStop}{bernevig06b}%
\bibitem{kane05}%
  \BibitemOpen
  \bibfield{author}{%
  \bibinfo {author} {\bibfnamefont{C.~L.}\ \bibnamefont{Kane}}\ and\ \bibinfo
  {author} {\bibfnamefont{E.~J.}\ \bibnamefont{Mele}},\ }%
  \bibfield{journal}{%
  \bibinfo {journal} {Phys. Rev. Lett.}\ }%
  \textbf{\bibinfo {volume} {95}},\ \bibinfo {pages} {226801} (\bibinfo {year}
  {2005})%
  \bibAnnoteFile{NoStop}{kane05}%
\bibitem{xu06}%
  \BibitemOpen
  \bibfield{author}{%
  \bibinfo {author} {\bibfnamefont{C.}~\bibnamefont{Xu}}\ and\ \bibinfo
  {author} {\bibfnamefont{J.~E.}\ \bibnamefont{Moore}},\ }%
  \bibfield{journal}{%
  \bibinfo {journal} {Phys. Rev. B}\ }%
  \textbf{\bibinfo {volume} {73}},\ \bibinfo {pages} {045322} (\bibinfo {year}
  {2006})%
  \bibAnnoteFile{NoStop}{xu06}%
\bibitem{wu06}%
  \BibitemOpen
  \bibfield{author}{%
  \bibinfo {author} {\bibfnamefont{C.}~\bibnamefont{Wu}}, \bibinfo {author}
  {\bibfnamefont{B.~A.}\ \bibnamefont{Bernevig}},\ and\ \bibinfo {author}
  {\bibfnamefont{S.}~\bibnamefont{Zhang}},\ }%
  \bibfield{journal}{%
  \bibinfo {journal} {Phys. Rev. Lett.}\ }%
  \textbf{\bibinfo {volume} {96}},\ \bibinfo {pages} {106401} (\bibinfo {year}
  {2006})%
  \bibAnnoteFile{NoStop}{wu06}%
\bibitem{maciejko09}%
  \BibitemOpen
  \bibfield{author}{%
  \bibinfo {author} {\bibfnamefont{J.}~\bibnamefont{Maciejko}}, \bibinfo
  {author} {\bibfnamefont{C.}~\bibnamefont{Liu}}, \bibinfo {author}
  {\bibfnamefont{Y.}~\bibnamefont{Oreg}}, \bibinfo {author}
  {\bibfnamefont{X.}~\bibnamefont{Qi}}, \bibinfo {author}
  {\bibfnamefont{C.}~\bibnamefont{Wu}},\ and\ \bibinfo {author}
  {\bibfnamefont{S.}~\bibnamefont{Zhang}},\ }%
  \bibfield{journal}{%
  \bibinfo {journal} {Phys. Rev. Lett.}\ }%
  \textbf{\bibinfo {volume} {102}},\ \bibinfo {pages} {256803} (\bibinfo {year}
  {2009})%
  \bibAnnoteFile{NoStop}{maciejko09}%
\bibitem{delplace12}%
  \BibitemOpen
  \bibfield{author}{%
  \bibinfo {author} {\bibfnamefont{P.}~\bibnamefont{Delplace}}, \bibinfo
  {author} {\bibfnamefont{J.}~\bibnamefont{Li}},\ and\ \bibinfo {author}
  {\bibfnamefont{M.}~\bibnamefont{B\"uttiker}},\ }%
  \bibfield{journal}{%
  \bibinfo {journal} {Phys. Rev. Lett.}\ }%
  \textbf{\bibinfo {volume} {109}},\ \bibinfo {pages} {246803} (\bibinfo {year}
  {2012})%
  \bibAnnoteFile{NoStop}{delplace12}%
\bibitem{zhou08}%
  \BibitemOpen
  \bibfield{author}{%
  \bibinfo {author} {\bibfnamefont{B.}~\bibnamefont{Zhou}}, \bibinfo {author}
  {\bibfnamefont{H.-Z.}\ \bibnamefont{Lu}}, \bibinfo {author}
  {\bibfnamefont{R.-L.}\ \bibnamefont{Chu}}, \bibinfo {author}
  {\bibfnamefont{S.-Q.}\ \bibnamefont{Shen}},\ and\ \bibinfo {author}
  {\bibfnamefont{Q.}~\bibnamefont{Niu}},\ }%
  \bibfield{journal}{%
  \bibinfo {journal} {Phys. Rev. Lett.}\ }%
  \textbf{\bibinfo {volume} {101}},\ \bibinfo {pages} {246807} (\bibinfo {year}
  {2008})%
  \bibAnnoteFile{NoStop}{zhou08}%
\bibitem{strom09}%
  \BibitemOpen
  \bibfield{author}{%
  \bibinfo {author} {\bibfnamefont{A.}~\bibnamefont{Str\"om}}\ and\ \bibinfo
  {author} {\bibfnamefont{H.}~\bibnamefont{Johannesson}},\ }%
  \bibfield{journal}{%
  \bibinfo {journal} {Phys. Rev. Lett.}\ }%
  \textbf{\bibinfo {volume} {102}},\ \bibinfo {pages} {096806} (\bibinfo {year}
  {2009})%
  \bibAnnoteFile{NoStop}{strom09}%
\bibitem{hou09}%
  \BibitemOpen
  \bibfield{author}{%
  \bibinfo {author} {\bibfnamefont{C.-Y.}\ \bibnamefont{Hou}}, \bibinfo
  {author} {\bibfnamefont{E.-A.}\ \bibnamefont{Kim}},\ and\ \bibinfo {author}
  {\bibfnamefont{C.}~\bibnamefont{Chamon}},\ }%
  \bibfield{journal}{%
  \bibinfo {journal} {Phys. Rev. Lett.}\ }%
  \textbf{\bibinfo {volume} {102}},\ \bibinfo {pages} {076602} (\bibinfo {year}
  {2009})%
  \bibAnnoteFile{NoStop}{hou09}%
\bibitem{teo09}%
  \BibitemOpen
  \bibfield{author}{%
  \bibinfo {author} {\bibfnamefont{J.~C.~Y.}\ \bibnamefont{Teo}}\ and\ \bibinfo
  {author} {\bibfnamefont{C.~L.}\ \bibnamefont{Kane}},\ }%
  \bibfield{journal}{%
  \bibinfo {journal} {Phys. Rev. B}\ }%
  \textbf{\bibinfo {volume} {79}},\ \bibinfo {pages} {235321} (\bibinfo {year}
  {2009})%
  \bibAnnoteFile{NoStop}{teo09}%
\bibitem{liu08}%
  \BibitemOpen
  \bibfield{author}{%
  \bibinfo {author} {\bibfnamefont{C.}~\bibnamefont{Liu}}, \bibinfo {author}
  {\bibfnamefont{T.~L.}\ \bibnamefont{Hughes}}, \bibinfo {author}
  {\bibfnamefont{X.-L.}\ \bibnamefont{Qi}}, \bibinfo {author}
  {\bibfnamefont{K.}~\bibnamefont{Wang}},\ and\ \bibinfo {author}
  {\bibfnamefont{S.-C.}\ \bibnamefont{Zhang}},\ }%
  \bibfield{journal}{%
  \bibinfo {journal} {Phys. Rev. Lett.}\ }%
  \textbf{\bibinfo {volume} {100}},\ \bibinfo {pages} {236601} (\bibinfo {year}
  {2008})%
  \bibAnnoteFile{NoStop}{liu08}%
\bibitem{liu11}%
  \BibitemOpen
  \bibfield{author}{%
  \bibinfo {author} {\bibfnamefont{C.-X.}\ \bibnamefont{Liu}}, \bibinfo
  {author} {\bibfnamefont{J.~C.}\ \bibnamefont{Budich}}, \bibinfo {author}
  {\bibfnamefont{P.}~\bibnamefont{Recher}},\ and\ \bibinfo {author}
  {\bibfnamefont{B.}~\bibnamefont{Trauzettel}},\ }%
  \bibfield{journal}{%
  \bibinfo {journal} {Phys. Rev. B}\ }%
  \textbf{\bibinfo {volume} {83}},\ \bibinfo {pages} {035407} (\bibinfo {year}
  {2011})%
  \bibAnnoteFile{NoStop}{liu11}%
\bibitem{schmidt11}%
  \BibitemOpen
  \bibfield{author}{%
  \bibinfo {author} {\bibfnamefont{T.~L.}\ \bibnamefont{Schmidt}},\ }%
  \bibfield{journal}{%
  \bibinfo {journal} {Phys. Rev. Lett.}\ }%
  \textbf{\bibinfo {volume} {107}},\ \bibinfo {pages} {096602} (\bibinfo {year}
  {2011})%
  \bibAnnoteFile{NoStop}{schmidt11}%
\bibitem{dolcini11}%
  \BibitemOpen
  \bibfield{author}{%
  \bibinfo {author} {\bibfnamefont{F.}~\bibnamefont{Dolcini}},\ }%
  \bibfield{journal}{%
  \bibinfo {journal} {Phys. Rev. B}\ }%
  \textbf{\bibinfo {volume} {83}},\ \bibinfo {pages} {165304} (\bibinfo {year}
  {2011})%
  \bibAnnoteFile{NoStop}{dolcini11}%
\bibitem{lee12}%
  \BibitemOpen
  \bibfield{author}{%
  \bibinfo {author} {\bibfnamefont{Y.-W.}\ \bibnamefont{Lee}}, \bibinfo
  {author} {\bibfnamefont{Y.-L.}\ \bibnamefont{Lee}},\ and\ \bibinfo {author}
  {\bibfnamefont{C.-H.}\ \bibnamefont{Chung}},\ }%
  \bibfield{journal}{%
  \bibinfo {journal} {Phys. Rev. B}\ }%
  \textbf{\bibinfo {volume} {86}},\ \bibinfo {pages} {235121} (\bibinfo {year}
  {2012})%
  \bibAnnoteFile{NoStop}{lee12}%
\bibitem{dolcetto12}%
  \BibitemOpen
  \bibfield{author}{%
  \bibinfo {author} {\bibfnamefont{G.}~\bibnamefont{Dolcetto}}, \bibinfo
  {author} {\bibfnamefont{S.}~\bibnamefont{Barbarino}}, \bibinfo {author}
  {\bibfnamefont{D.}~\bibnamefont{Ferraro}}, \bibinfo {author}
  {\bibfnamefont{N.}~\bibnamefont{Magnoli}},\ and\ \bibinfo {author}
  {\bibfnamefont{M.}~\bibnamefont{Sassetti}},\ }%
  \bibfield{journal}{%
  \bibinfo {journal} {Phys. Rev. B}\ }%
  \textbf{\bibinfo {volume} {85}},\ \bibinfo {pages} {195138} (\bibinfo {year}
  {2012})%
  \bibAnnoteFile{NoStop}{dolcetto12}%
\bibitem{edge13}%
  \BibitemOpen
  \bibfield{author}{%
  \bibinfo {author} {\bibfnamefont{J.~M.}\ \bibnamefont{Edge}}, \bibinfo
  {author} {\bibfnamefont{J.}~\bibnamefont{Li}}, \bibinfo {author}
  {\bibfnamefont{P.}~\bibnamefont{Delplace}},\ and\ \bibinfo {author}
  {\bibfnamefont{M.}~\bibnamefont{B\"uttiker}},\ }%
  \bibfield{journal}{%
  \bibinfo {journal} {Phys. Rev. Lett.}\ }%
  \textbf{\bibinfo {volume} {110}},\ \bibinfo {pages} {246601} (\bibinfo {year}
  {2013})%
  \bibAnnoteFile{NoStop}{edge13}%
\bibitem{skinner12}%
  \BibitemOpen
  \bibfield{author}{%
  \bibinfo {author} {\bibfnamefont{B.}~\bibnamefont{Skinner}}, \bibinfo
  {author} {\bibfnamefont{T.}~\bibnamefont{Chen}},\ and\ \bibinfo {author}
  {\bibfnamefont{B.~I.}\ \bibnamefont{Shklovskii}},\ }%
  \bibfield{journal}{%
  \bibinfo {journal} {Phys. Rev. Lett.}\ }%
  \textbf{\bibinfo {volume} {109}},\ \bibinfo {pages} {176801} (\bibinfo {year}
  {2012})%
  \bibAnnoteFile{NoStop}{skinner12}%
\bibitem{vayrynen13}%
  \BibitemOpen
  \bibfield{author}{%
  \bibinfo {author} {\bibfnamefont{J.~I.}\ \bibnamefont{V\"ayrynen}}, \bibinfo
  {author} {\bibfnamefont{M.}~\bibnamefont{Goldstein}},\ and\ \bibinfo {author}
  {\bibfnamefont{L.~I.}\ \bibnamefont{Glazman}},\ }%
  \bibfield{journal}{%
  \bibinfo {journal} {Phys. Rev. Lett.}\ }%
  \textbf{\bibinfo {volume} {110}},\ \bibinfo {pages} {216402} (\bibinfo {year}
  {2013})%
  \bibAnnoteFile{NoStop}{vayrynen13}%
\bibitem{schmidt12}%
  \BibitemOpen
  \bibfield{author}{%
  \bibinfo {author} {\bibfnamefont{T.~L.}\ \bibnamefont{Schmidt}}, \bibinfo
  {author} {\bibfnamefont{S.}~\bibnamefont{Rachel}}, \bibinfo {author}
  {\bibfnamefont{F.}~\bibnamefont{von Oppen}},\ and\ \bibinfo {author}
  {\bibfnamefont{L.~I.}\ \bibnamefont{Glazman}},\ }%
  \bibfield{journal}{%
  \bibinfo {journal} {Phys. Rev. Lett.}\ }%
  \textbf{\bibinfo {volume} {108}},\ \bibinfo {pages} {156402} (\bibinfo {year}
  {2012})%
  \bibAnnoteFile{NoStop}{schmidt12}%
\bibitem{rothe10}%
  \BibitemOpen
  \bibfield{author}{%
  \bibinfo {author} {\bibfnamefont{D.~G.}\ \bibnamefont{Rothe}}, \bibinfo
  {author} {\bibfnamefont{R.~W.}\ \bibnamefont{Reinthaler}}, \bibinfo {author}
  {\bibfnamefont{C.}~\bibnamefont{Liu}}, \bibinfo {author}
  {\bibfnamefont{L.~W.}\ \bibnamefont{Molenkamp}}, \bibinfo {author}
  {\bibfnamefont{S.}~\bibnamefont{Zhang}},\ and\ \bibinfo {author}
  {\bibfnamefont{E.~M.}\ \bibnamefont{Hankiewicz}},\ }%
  \bibfield{journal}{%
  \bibinfo {journal} {New J. Phys.}\ }%
  \textbf{\bibinfo {volume} {12}},\ \bibinfo {pages} {065012} (\bibinfo {year}
  {2010})%
  \bibAnnoteFile{NoStop}{rothe10}%
\bibitem{bernevig06}%
  \BibitemOpen
  \bibfield{author}{%
  \bibinfo {author} {\bibfnamefont{B.~A.}\ \bibnamefont{Bernevig}}, \bibinfo
  {author} {\bibfnamefont{T.~L.}\ \bibnamefont{Hughes}},\ and\ \bibinfo
  {author} {\bibfnamefont{S.-C.}\ \bibnamefont{Zhang}},\ }%
  \bibfield{journal}{%
  \bibinfo {journal} {Science}\ }%
  \textbf{\bibinfo {volume} {314}},\ \bibinfo {pages} {1757} (\bibinfo {year}
  {2006})%
  \bibAnnoteFile{NoStop}{bernevig06}%
\bibitem{knez11}%
  \BibitemOpen
  \bibfield{author}{%
  \bibinfo {author} {\bibfnamefont{I.}~\bibnamefont{Knez}}, \bibinfo {author}
  {\bibfnamefont{R.-R.}\ \bibnamefont{Du}},\ and\ \bibinfo {author}
  {\bibfnamefont{G.}~\bibnamefont{Sullivan}},\ }%
  \bibfield{journal}{%
  \bibinfo {journal} {Phys. Rev. Lett.}\ }%
  \textbf{\bibinfo {volume} {107}},\ \bibinfo {pages} {136603} (\bibinfo {year}
  {2011})%
  \bibAnnoteFile{NoStop}{knez11}%
\bibitem{chu09}%
  \BibitemOpen
  \bibfield{author}{%
  \bibinfo {author} {\bibfnamefont{R.-L.}\ \bibnamefont{Chu}}, \bibinfo
  {author} {\bibfnamefont{J.}~\bibnamefont{Li}}, \bibinfo {author}
  {\bibfnamefont{J.~K.}\ \bibnamefont{Jain}},\ and\ \bibinfo {author}
  {\bibfnamefont{S.-Q.}\ \bibnamefont{Shen}},\ }%
  \bibfield{journal}{%
  \bibinfo {journal} {Phys. Rev. B}\ }%
  \textbf{\bibinfo {volume} {80}},\ \bibinfo {pages} {081102} (\bibinfo {year}
  {2009})%
  \bibAnnoteFile{NoStop}{chu09}%
\bibitem{virtanen11}%
  \BibitemOpen
  \bibfield{author}{%
  \bibinfo {author} {\bibfnamefont{P.}~\bibnamefont{Virtanen}}\ and\ \bibinfo
  {author} {\bibfnamefont{P.}~\bibnamefont{Recher}},\ }%
  \bibfield{journal}{%
  \bibinfo {journal} {Phys. Rev. B}\ }%
  \textbf{\bibinfo {volume} {83}},\ \bibinfo {pages} {115332} (\bibinfo {year}
  {2011})%
  \bibAnnoteFile{NoStop}{virtanen11}%
\bibitem{romeo12}%
  \BibitemOpen
  \bibfield{author}{%
  \bibinfo {author} {\bibfnamefont{F.}~\bibnamefont{Romeo}}, \bibinfo {author}
  {\bibfnamefont{R.}~\bibnamefont{Citro}}, \bibinfo {author}
  {\bibfnamefont{D.}~\bibnamefont{Ferraro}},\ and\ \bibinfo {author}
  {\bibfnamefont{M.}~\bibnamefont{Sassetti}},\ }%
  \bibfield{journal}{%
  \bibinfo {journal} {Phys. Rev. B}\ }%
  \textbf{\bibinfo {volume} {86}},\ \bibinfo {pages} {165418} (\bibinfo {year}
  {2012})%
  \bibAnnoteFile{NoStop}{romeo12}%
\bibitem{koenig13}%
  \BibitemOpen
  \bibfield{author}{%
  \bibinfo {author} {\bibfnamefont{M.}~\bibnamefont{K\"onig}}, \bibinfo
  {author} {\bibfnamefont{M.}~\bibnamefont{Baenninger}}, \bibinfo {author}
  {\bibfnamefont{A.~G.~F.}\ \bibnamefont{Garcia}}, \bibinfo {author}
  {\bibfnamefont{N.}~\bibnamefont{Harjee}}, \bibinfo {author}
  {\bibfnamefont{B.~L.}\ \bibnamefont{Pruitt}}, \bibinfo {author}
  {\bibfnamefont{C.}~\bibnamefont{Ames}}, \bibinfo {author}
  {\bibfnamefont{P.}~\bibnamefont{Leubner}}, \bibinfo {author}
  {\bibfnamefont{C.}~\bibnamefont{Br\"une}}, \bibinfo {author}
  {\bibfnamefont{H.}~\bibnamefont{Buhmann}}, \bibinfo {author}
  {\bibfnamefont{L.~W.}\ \bibnamefont{Molenkamp}},\ and\ \bibinfo {author}
  {\bibfnamefont{D.}~\bibnamefont{Goldhaber-Gordon}},\ }%
  \bibfield{journal}{%
  \bibinfo {journal} {Phys. Rev. X}\ }%
  \textbf{\bibinfo {volume} {3}},\ \bibinfo {pages} {021003} (\bibinfo {year}
  {2013})%
  \bibAnnoteFile{NoStop}{koenig13}%
\bibitem{bruus04}%
  \BibitemOpen
  \bibfield{author}{%
  \bibinfo {author} {\bibfnamefont{H.}~\bibnamefont{Bruus}}\ and\ \bibinfo
  {author} {\bibfnamefont{K.}~\bibnamefont{Flensberg}},\ }%
  \emph{\bibinfo {title} {Many-Body Quantum Theory in Condensed Matter
  Physics}}\ (\bibinfo {publisher} {Oxford University Press},\ \bibinfo {year}
  {2004})%
  \bibAnnoteFile{NoStop}{bruus04}%
\bibitem{anderson80}%
  \BibitemOpen
  \bibfield{author}{%
  \bibinfo {author} {\bibfnamefont{P.~W.}\ \bibnamefont{Anderson}}, \bibinfo
  {author} {\bibfnamefont{D.~J.}\ \bibnamefont{Thouless}}, \bibinfo {author}
  {\bibfnamefont{E.}~\bibnamefont{Abrahams}},\ and\ \bibinfo {author}
  {\bibfnamefont{D.~S.}\ \bibnamefont{Fisher}},\ }%
  \bibfield{journal}{%
  \bibinfo {journal} {Phys. Rev. B}\ }%
  \textbf{\bibinfo {volume} {22}},\ \bibinfo {pages} {3519} (\bibinfo {year}
  {1980})%
  \bibAnnoteFile{NoStop}{anderson80}%
\bibitem{pendry94}%
  \BibitemOpen
  \bibfield{author}{%
  \bibinfo {author} {\bibfnamefont{J.}~\bibnamefont{Pendry}},\ }%
  \bibfield{journal}{%
  \bibinfo {journal} {Advances in Physics}\ }%
  \textbf{\bibinfo {volume} {43}},\ \bibinfo {pages} {461} (\bibinfo {year}
  {1994})%
  \bibAnnoteFile{NoStop}{pendry94}%
\end{thebibliography}%

\end{document}